\documentclass[aps,twocolumn,showpacs,preprintnumbers,amsmath,amssymb,nofootinbib]{revtex4}
\usepackage{array}
\usepackage{epsfig,float}
\usepackage{amsmath,amsfonts,amssymb,graphicx,bm,bbm}
\usepackage{dcolumn}

\providecommand{\tabularnewline}{\\}

\def\u1x{${U(1)_X}$}
\def\beq{\begin{equation}}
\def\eeq{\end{equation}}
\def\bea{\begin{eqnarray}}
\def\eea{\end{eqnarray}}
\def\bmat{\begin{pmatrix}}
\def\emat{\end{pmatrix}}

\def\mev{\,{\rm MeV}}
\def\gev{\,{\rm GeV}}

\def\zpri{{Z^{\prime}}}
\def\mzpri{{M_{Z^\prime}}}

\begin{document}

\preprint{CERN-PH-TH/2008-11;\ MCTP-07-47 ;\ BNL-HET-08/2}

\title{Higgs boson decays to four fermions through an abelian hidden sector}

%
%

%
\author{Shrihari Gopalakrishna$^{a}$}
\author{Sunghoon Jung$^{b}$}
\author{James D. Wells$^{c,b}$}

\vspace{0.2cm}

\affiliation{
${}^a$ Physics Department, Brookhaven National Laboratory, Upton, NY 11973 \\
${}^b$ MCTP, University of Michigan, Ann Arbor, MI 48109 \\
${}^c$ CERN, Theory Division (PH-TH), CH-1211 Geneva 23, Switzerland }



\begin{abstract}

We consider a generic abelian hidden sector that couples to the Standard Model only through gauge-invariant renormalizable operators.  This allows the exotic Higgs boson to mix with the Standard Model Higgs boson, and the exotic abelian gauge boson to mix with the Standard Model hypercharge gauge boson. One immediate consequence of spontaneous breaking of the hidden sector gauge group is the possible decay of the lightest Higgs boson into four fermions through intermediate exotic gauge bosons. We study the implications of this decay for Higgs boson phenomenology at the Fermilab Tevatron Collider and the CERN Large Hadron Collider. Our emphasis is on the four lepton final state.

\end{abstract}


\maketitle


{\it Introduction.}
The fundamental theory may be significantly richer than the Standard Model (SM) world that
we have directly probed.  Copies of many other gauge theories may be inaccessible to us because the particles that form our bodies are not charged under them. Is there a method to explore such
hidden worlds given the limited collection of charges that we can directly probe?  The answer is not assured, but an opportunity can be identified~\cite{Schabinger:2005ei,Patt:2006fw,Strassler:2006im,Kumar:2006gm,John:today}.

The SM has two gauge invariant, flavor-neutral operators that are relevant (dimension $<4$): the hypercharge field-strength tensor $B_{\mu\nu}$ and
the SM Higgs mass operator $|\Phi_{SM}|^2$.  Hidden sector (i.e., non-SM states with no SM charge) abelian gauge bosons $X$ and Higgs bosons $\Phi_H$ can couple to these operators in a gauge invariant, renormalizable manner~\cite{third possibility}:
\beq
X_{\mu\nu}B^{\mu\nu},~~~{\rm and}~~~|\Phi_H|^2|\Phi_{SM}|^2.
\label{renormalizable ops}
\eeq
These couplings give us the opportunity we are looking for to see the effects of a hidden sector by virtue of their interactions with states we can observe.

In this letter we investigate the implications for Higgs boson phenomenology of the simultaneous existence of the two operators in Eq.~(\ref{renormalizable ops}).  We do not tie ourselves to any particular model of the hidden abelian sector. We note that if the kinetic mixing between the gauge bosons is large, precision electroweak
 and dedicated collider searches may see the
 effects~\cite{Babu:1997st, Chang:2006fp,Kumar:2006gm,Rizzo:2006nw,Langacker:2008yv}.
 For our purposes, we only need the kinetic mixing to be non-zero and large enough to allow prompt decays of the exotic gauge boson eigenstate.   We also note that the pure mixing effects of $\Phi_H$ and $\Phi_{SM}$ can be probed well by
 colliders~\cite{Schabinger:2005ei,Strassler:2006im,Barbieri:2005ri,Chacko:2005vw,Bowen:2007ia}
   even if no exotic decay modes are kinematically accessible. However, it would be more difficult in that circumstance to know what the origin is of the shift in Higgs boson phenomenology at colliders.
For related discussion on the phenomenology of a hidden sector see
Ref.~\cite{Foot:1991bp}.

Instead, what we focus on here is the prospect of the exotic gauge boson being sufficiently light such that the lightest Higgs boson decays into a pair of them~\cite{Strassler:2008bv}.  The decay of the Higgs boson into two $X$ bosons is through Higgs boson mixing.  The $X$ boson will then decay into SM fermions if there is even a tiny amount of kinetic mixing, which we assume to be the case.
The $X$ bosons could have competing branching fraction
into other exotic states potentially leading to invisible decays or even more background-free
topologies than considered here. We neglect these possibilities in order to keep our analysis simple
and our assumptions to a minimum.
We are particularly interested in leptonic final states.
Thus, the subject of this paper is to provide the details of how $pp\to h\to XX\to \bar l l \bar l' l'$ is possible within this theoretical framework, and to explore the detectability of this channel
at the Fermilab Tevatron and CERN LHC.


{\it Theory framework.}
We consider an extra $U(1)_X$ factor in addition to the SM gauge group.
The only coupling of this new gauge sector to the SM is through kinetic mixing with
the hypercharge gauge boson $B_\mu$.
The kinetic energy terms of the \u1x gauge group are
\beq
{\cal L}^{KE}_X = -\frac{1}{4} \hat{X}_{\mu\nu} \hat{X}^{\mu\nu} + \frac{\chi}{2} \hat{X}_{\mu\nu} \hat{B}^{\mu\nu} \ ,
\eeq
where we take the parameter $\chi \ll 1$ to be consistent with precision electroweak
constraints.  Hats on fields imply that gauge fields do not have canonically normalized kinetic terms.

As an example, we note that heavy states that are charged under both $U(1)_Y$ and $U(1)_X$
can typically induce a $\chi$ at the loop level~\cite{Holdom:1985ag} given by
$\chi \sim g^\prime g_X/(16\pi^2) \sim 10^{-3}$.
Tree-level mixing, although possible, will be absent if the $U(1)_X$ is the remnant of
a spontaneously broken non-abelian gauge symmetry.
If the scale of $U(1)_X$ breaking is not too far above the electroweak scale,
a radiatively generated $\chi$ will be quite small. We take
the $U(1)_X$ breaking VEV $\xi \sim 1$~TeV.

We introduce a new Higgs boson $\Phi_{H}$ in addition to the usual SM Higgs boson
$\Phi_{SM}$.
Under $SU(2)_L \otimes U(1)_Y \otimes U(1)_X$ we take the representations
$\Phi_{SM}: (2, 1/2, 0)$ and $\Phi_{H}: (1, 0, q_X)$, with $q_X$ arbitrary.
The Higgs sector Lagrangian is
\bea
{\cal L}_{\Phi} &=& |D_\mu \Phi_{SM}|^2
+ |D_\mu \Phi_H |^2 
 + m^2_{\Phi_H}|\Phi_H|^2 + m^2_{\Phi_{SM}}|\Phi_{SM}|^2 \nonumber\\
& & - \lambda|\Phi_{SM}|^4 - \rho|\Phi_H|^4 - \kappa
|\Phi_{SM}|^2|\Phi_H|^2. \label{Lphi.EQ}
\eea
so that $U(1)_X$ is broken spontaneously by $\left< \Phi_H \right> = \xi/\sqrt{2}$,
and electroweak symmetry is broken spontaneously as usual by
$\left< \Phi_{SM}\right> = (0,v/\sqrt{2})$.

One can diagonalize the kinetic terms by redefining $\hat{X}_\mu ,
\hat{B}_\mu \rightarrow X_\mu , B_\mu$ with
\begin{displaymath}
\left( \begin{array}{c} X_\mu \\ B_\mu \end{array} \right) = \left(
\begin{array}{cc} \sqrt{1-\chi^2} & 0 \\ -\chi & 1 \end{array} \right)
\left( \begin{array}{c} \hat{X}_\mu \\ \hat{B}_\mu \end{array}
\right) \label{higgs.eq}
\end{displaymath}
The covariant derivative is then\beq D_\mu =
\partial_\mu + i (g_X Q_X + g^\prime \eta Q_Y) X_\mu + i g^\prime
Q_Y B_\mu + i g T^3 W^3_\mu \  . \label{DMUGAU.EQ} \eeq
where $\eta \equiv \chi / \sqrt{1-\chi^2}$.

After a $GL(2,R)$ rotation to diagonalize the kinetic terms followed by an $O(3)$ rotation
to diagonalize the $3 \times 3$ neutral gauge boson mass matrix, we can write the
mass eigenstates as (with $s_x\equiv \sin{\theta_x}$,  $c_x\equiv \cos{\theta_x}$)
\bea
\begin{pmatrix} B \\ W^3 \\ X  \end{pmatrix} =
\begin{pmatrix}
c_W & -s_W c_\alpha  & s_W s_\alpha \\
s_W & c_W c_\alpha & -c_W s_\alpha \\
0 & s_\alpha & c_\alpha
\end{pmatrix}
\begin{pmatrix} A \\ Z \\ \zpri  \end{pmatrix} \ ,
\label{GAU2MAS.EQ}
\eea
where the
usual weak mixing angle and the new gauge boson mixing angle  are
\beq
s_W \equiv \frac{g^\prime}{\sqrt{g^2 + {g^\prime}^2}} \ ; \quad
\tan{\left( 2\theta_\alpha \right)} = \frac{-2 s_W \eta}{1 - s_W^2\eta^2 - \Delta_Z} \ ,
\label{tan2al.EQ}
\eeq
with $\Delta_Z =
M_{X}^2/M_{Z_0}^2$, $M_{X}^2 = \xi^2 g_X^2 q_X^2$, $M_{Z_0}^2 =
(g^2 + {g^\prime}^2) v^2 / 4$. $M_{Z_0}$ and $M_X$ are masses before mixing. The photon is massless (i.e., $M_A = 0 $), and the two heavier gauge boson
mass eigenvalues are
\begin{eqnarray} M_{Z, Z^\prime} &=& \frac{M_{Z_0}^2}{2}
\left[\left(1+s_W^2 \eta^2 + \Delta_Z \right) \right. \nonumber\\ &&
\left. \qquad \pm \sqrt{\left( 1 - s_W^2 \eta^2 - \Delta_Z \right)^2
+ 4 s_W^2 \eta^2 } \right] ,
\end{eqnarray}
valid for $\Delta_Z < (1-s_W^2 \eta^2)$ ($Z \leftrightarrow Z^\prime$ otherwise). Since we assume that $\eta \ll 1$, mass
eigenvalues are taken as $M_Z \approx M_{Z_0}=91.19$ GeV and
$M_\zpri \approx M_X$.

The two real physical Higgs bosons $\phi_{SM}$ and $\phi_H$ mix after symmetry breaking,
and the mass eigenstates $h, H$ are
\begin{displaymath} \left( \begin{array}{c} \phi_{SM} \\ \phi_H
\end{array} \right) = \left( \begin{array}{cc} c_h & s_h \\ -s_h &
c_h \end{array} \right) \left( \begin{array}{c} h \\ H \end{array}
\right).
\end{displaymath}
The mixing angle and mass eigenvalues are
\begin{eqnarray}
\tan{(2\theta_h)} &=& \frac{\kappa v \xi}{\rho \xi^2 - \lambda v^2}
\
\\ M_{h,H}^2 = \left( \lambda v^2 + \rho \xi^2 \right)
            &\mp& \sqrt{ (\lambda v^2 - \rho \xi^2)^2 + \kappa^2 v^2 \xi^2} \ .
\end{eqnarray}
Although the mixing angle depends on the many unknown parameters of Eq.(\ref{Lphi.EQ}),
we will treat the resulting $\theta_h$ as an input along with the Higgs boson masses.

Now we are able to present the couplings of the $\zpri$ to various SM states.

\underline{Fermion couplings}:
Couplings to SM fermions are
\bea
\bar \psi\psi Z \!\!&:&\!\! \frac{ig}{c_W} \left[ c_\alpha ( 1 - s_W t_\alpha \eta ) \right]
\left[  T^3_L - \frac{(1 - t_\alpha \eta/s_W)}{( 1 - s_W t_\alpha \eta)}  s_W^2 Q  \right]
\nonumber \\
\bar\psi\psi\zpri\!\!\!\! &:&\!\!\! \frac{-ig}{c_W} \left[c_\alpha  (t_\alpha + \eta s_W) \right] \left[ T^3_L - \frac{ (t_\alpha + \eta/s_W)}{ (t_\alpha + \eta s_W)} s_W^2 Q \right]
\label{GBFerm.EQ}
\eea
where we have used $Q = T^3_L + Q_Y$ and $t_\alpha \equiv s_\alpha / c_\alpha$.
The photon coupling is as in the SM and is not shifted.

\underline{Triple gauge boson couplings}:
Denoting the coupling relative to the corresponding SM coupling as ${\cal R}$, we find
${\cal R}_{A W^+ W^-} = 1$, ${\cal R}_{Z W^+ W^-} = c_\alpha$ and
${\cal R}_{\zpri W^+ W^-} = - s_\alpha$ (the last is compared to the SM $Z W^+ W^-$ coupling).
In our case, to leading order we have $c_\alpha \approx 1$, $s_\alpha \ll 1$.

\underline{Higgs couplings}:
The Higgs couplings are
\beq
\label{hZpZ_coup.EQ}
\begin{split}
 h f f &:
-i c_h \frac{m_f}{v} \ , \qquad
h W W :
2i c_h \frac{M^2_W}{v} \, \\
h Z Z &:
2i c_h \frac{M^2_{Z_0}}{v} (-c_\alpha + \eta s_W s_\alpha)^2 - 2i s_h
\frac{M^2_X}{\xi} s^2_\alpha \ , \\
h \zpri \zpri &:
2i c_h \frac{M^2_{Z_0}}{v} (s_\alpha + \eta s_W c_\alpha)^2 -
2i s_h \frac{M^2_X}{\xi} c^2_\alpha \ , \\
h \zpri Z &:
2i c_h \frac{M^2_{Z_0}}{v} (-c_\alpha + \eta s_W s_\alpha)
(s_\alpha + \eta s_W c_\alpha) \\
&\qquad\qquad\qquad\qquad -2i s_h \frac{M^2_X}{\xi} s_\alpha c_\alpha \ .
\end{split}
\eeq

{\it Parameters and Precision Electroweak Constraints.}
Electroweak precision observables such as $M_W$, $\Gamma_Z$ and $A_{LR}$
constrain the  theory.
These constraints have been discussed in
greater detail in
Refs.~\cite{Babu:1997st,Kumar:2006gm,Chang:2006fp,Rizzo:2006nw,Langacker:2008yv}.
For our theory, given the experimental accuracy~\cite{Yao:2006px}
of precision electroweak observables including those mentioned above, we find the constraint
\beq
\frac{\eta}{\sqrt{|1-M_{Z'}^2/ M_Z^2 |}} \lesssim 10^{-2} \ .
\eeq
This is expected given
that the fractional accuracy of EW precision measurements are at the
$10^{-4}$ level, and in our model the deviations appear at ${\cal O}(\eta^2)$.


Fits to electroweak precision observables~\cite{Alcaraz:2006mx} constrain the
SM Higgs mass to be
$\log\left( M_{\rm Higgs}/1~{\rm GeV} \right) = 1.93^{+0.16}_{-0.17} $.
This can be turned into a constraint on our model by noting that all couplings to SM fields
involving $h$ have an additional factor of $c_h$ while those for $H$ have $s_h$, which results in
\beq
c_h^2 \log\left( \frac{M_h}{1~{\rm GeV}} \right) +  s_h^2 \log\left( \frac{M_H}{1~{\rm GeV}} \right)
\simeq 1.93^{+0.16}_{-0.17} \ .
\label{mhmH_fit.EQ}
\eeq
Equivalently, one can state the constraints in terms of the $S$ and $T$ parameters,
following the discussion in Ref.~\cite{Peskin:2001rw}. Since we do not specify the value of
the heavier Higgs mass, we have the freedom to choose it such that there is minimal difficulty
with precision electroweak constraints. Even if we choose a much heavier Higgs boson for
our second eigenstate, there are well-known ways the theory can be augmented to
be compatible with the data~\cite{Peskin:2001rw}.


{\it Decay Branching Fractions.}
We now turn to the actual decay branching fractions of the Higgs boson and $Z'$ mass eigenstate. We are particularly interested in the frequency of $h\to Z'Z'$ and the leptonic branching fractions of $Z'$.

\begin{figure}[t]
\begin{center}
\scalebox{0.48}{\includegraphics[angle=0]{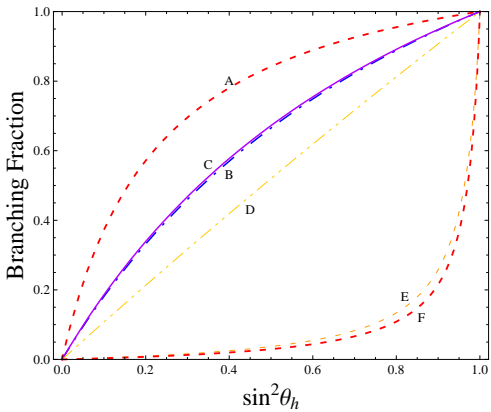}}
\caption{
Branching ratio of $h\to \zpri \zpri$ as a function of $s_h^2$ for various $\mzpri$
and $M_h$,
with $\eta = 10^{-4}$. Benchmark points are shown in Table \ref{bchmrk.TBL}.
\label{BRh2zpzp.FIG}
}
\end{center}
\end{figure}

\underline{$h\to \zpri\zpri$ decay}:
In Fig.~\ref{BRh2zpzp.FIG} we show the $h\to \zpri \zpri$ branching ratio
as a function of $s_h^2$, computed using HDECAY~\cite{Djouadi:1997yw}.
A $120\gev$ ($250\gev$) Higgs boson has total width of $\sim 10\mev$ ($\sim 2.1\gev$) when
$M_{Z'}=5\gev$ and $c_h^2=0.5$.
We do not include any heavy exotic
states that the $X_\mu$ couples to, which would require either considering the additional invisible decay  modes, studied well elsewhere, or much more spectacular and model-dependent
decay chains to SM particles.

%
%

\begin{figure}[t]
\begin{center}
\scalebox{0.45}{\includegraphics[angle=0,width=\textwidth]{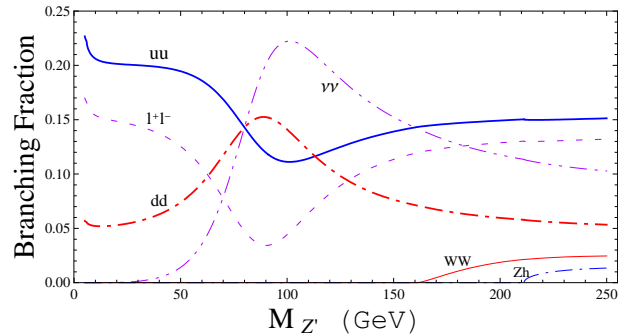}}
\caption{
Branching ratio of $\zpri$ into two body final states as a function of
$\mzpri$ with $c_h^2=0.5$ and $\eta=10^{-4}$.
\label{BRzp2yy.FIG}}
\end{center}
\end{figure}

\underline{$\zpri$ decay}:
 The $\zpri$ coupling to the SM sector is proportional to the tiny $\eta$,
making the width rather small, but these are the only modes kinematically
allowed for the $\zpri$ to decay into.
The $\zpri$ total width for $\eta=10^{-4}$ is $5.8\times 10^{-10}$, $2.7\times 10^{-9}$, $8.2\times 10^{-9}$ and $2.0\times 10^{-7}\gev$ for $M_{Z'}=5$, $20$, $50$ and $100\gev$ respectively.
This decay width is too small to be resolved by LHC experiment, but large enough to yield prompt decays.
The total width for any other $\eta$ can be obtained by scaling the above width
by $\eta^2$. Displaced vertices begin to be allowed when $\eta<10^{-5}$, which would be another interesting sign of exotic physics in the Higgs boson decays.
In Fig.~\ref{BRzp2yy.FIG} we show the $\zpri$ branching ratio into two body final states
as a function of $\mzpri$.

%

{\it Four Lepton Modes at the Tevatron and LHC.}
We focus on the mode $h\to \zpri\zpri\to 4\ell$ in our analysis
with $\ell = e,\mu$.
In presenting results in this section, we will choose
$\eta = 10^{-4}$, $\xi = 1$~TeV, and unless mentioned otherwise,
take $c_h^2 = 0.5$.
For illustration, we choose six benchmark points as shown in Table~\ref{bchmrk.TBL}
for which we compute the differential distributions,
make cuts and find the significance at the Tevatron and LHC.
We make use of the narrow width approximation and analyze in succession: $pp\to h$
followed by $h\to \zpri\zpri$ followed by $\zpri\to \ell^+\ell^-$.

\begin{table}
\begin{tabular}[c]{|c|c|c|c|c|c|c|}
\hline
Point & A & B & C & D & E & F \\
\hline
($M_h$, $\mzpri$) (GeV) & 120, 5 & 120, 50 & 150, 5 & 150, 50 & 250, 5 & 250, 50 \\
\hline
\end{tabular}
\caption{Six benchmark points that we study. \label{bchmrk.TBL}}
\end{table}

The gluon fusion process $gg\to h$ is the largest production channel at the
Tevatron ($\sqrt{s}=1.96$~TeV) and the LHC ($\sqrt{s}=14$~TeV).
For instance, at the Tevatron, NLO $\sigma(gg \rightarrow h) = 0.85$~pb
for $M_h=120$ GeV while the sum of the other channels gives $0.33$~pb; the corresponding
cross-sections at the LHC are $40.25$~pb and $7.7$~pb
respectively~\cite{Spira:1995mt, Djouadi:2005gi}. We include only gluon fusion computed at NLO
using  HIGLU~\cite{Spira:1995mt}.

The main sources of background are the SM processes
$pp \to h \to Z Z \to 4\ell$, and
$pp \to VV \to 4\ell$ where $VV$ denotes $ZZ$, $\gamma\gamma$ and $\gamma Z$ channels.
The $pp\to t\bar t $ production cross-section is large and 4-lepton events from
this process can be a source of (reducible) background at the LHC,
but we take it that this can be adequately suppressed (for details
see Ref.~\cite{atlas tdr}).

\begin{figure}[t]
\begin{center}
\includegraphics[angle=0,width=0.4\textwidth]{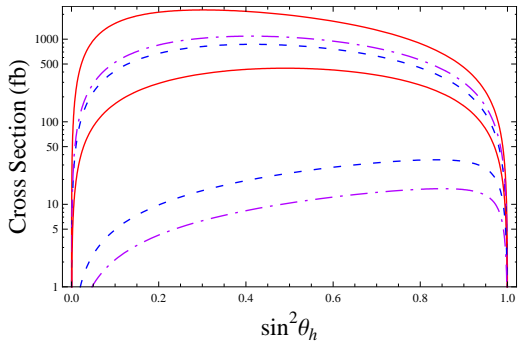}
\caption{Total cross section of the process $pp \rightarrow h
\rightarrow \zpri\zpri \rightarrow 4l$ at LHC as a function of
$\sin^2\theta_h$. From top to bottom, lines correspond to points
A,C,B,D,E,F. No cuts have been applied. \label{totcs.FIG}}
\end{center}
\end{figure}

We use MadGraph~\cite{Stelzer:1994ta} to obtain all matrix
elements, and generate event samples using MadEvent~\cite{Maltoni:2002qb}
with CTEQ6L1 PDF~\cite{Pumplin:2005rh}.
The cross-sections for the process $pp\to h\to \zpri\zpri\to 4\ell$
at the LHC without any cuts are shown in Fig.~\ref{totcs.FIG},
and the corresponding ones at the Tevatron while similar in shape,
are smaller by about 50, and will be discussed later in this section.
We present the $ee\mu\mu$ channels here, but this can be  extended to
include $4e$ and $4\mu$ channels.
The cross-section approaches zero as $c^2_h\to 1$ because $h$ will not couple to the
 $X$ boson, and also as $c_h^2\to 0$ because $h$ does not couple to the gluon in
this limit.
In these limits our analysis can be applied to probe the second Higgs
mass eigenstate $H$.

\begin{figure}[t]
\begin{center}
\includegraphics[angle=0,width=0.48\textwidth]{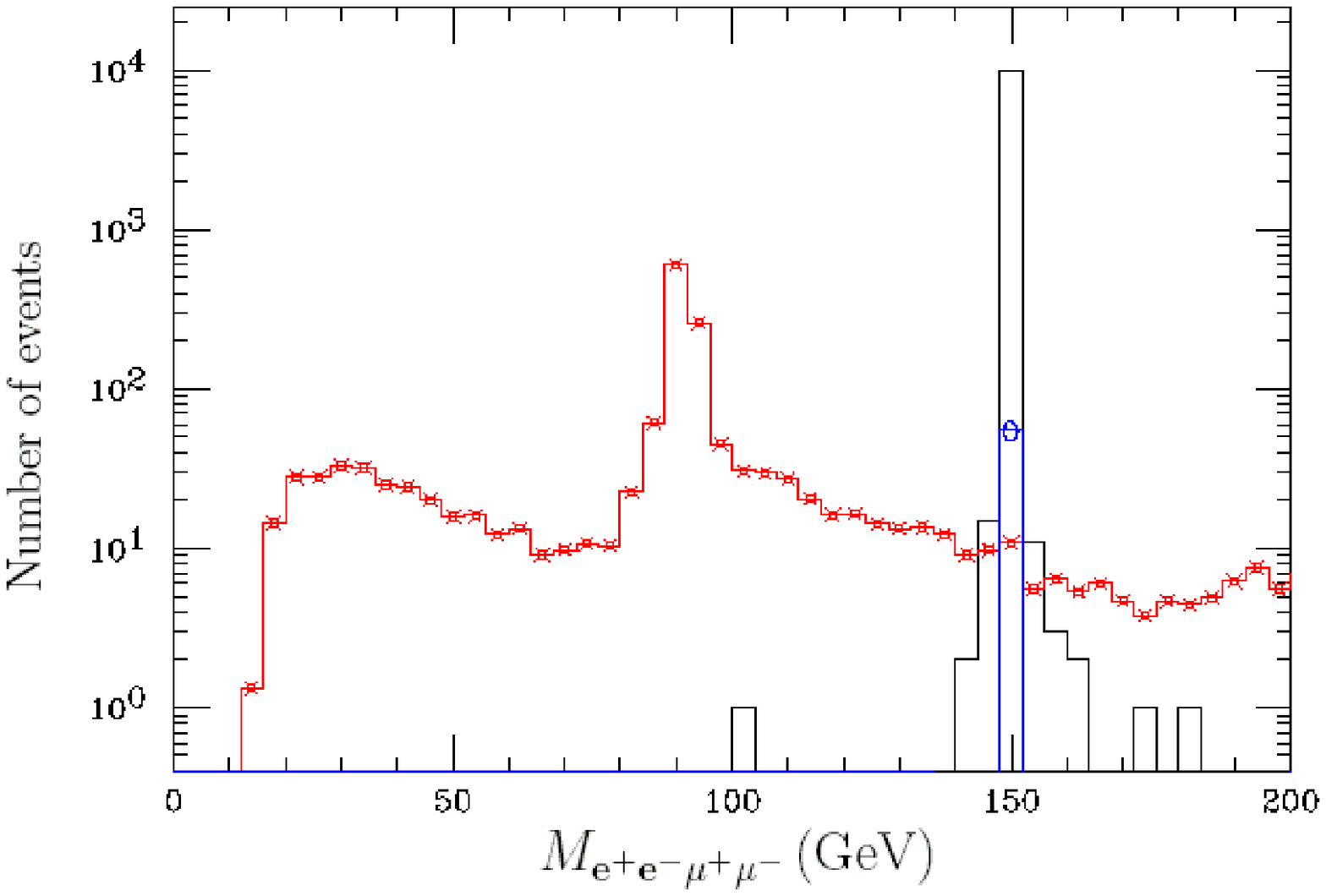}
\includegraphics[angle=0,width=0.48\textwidth]{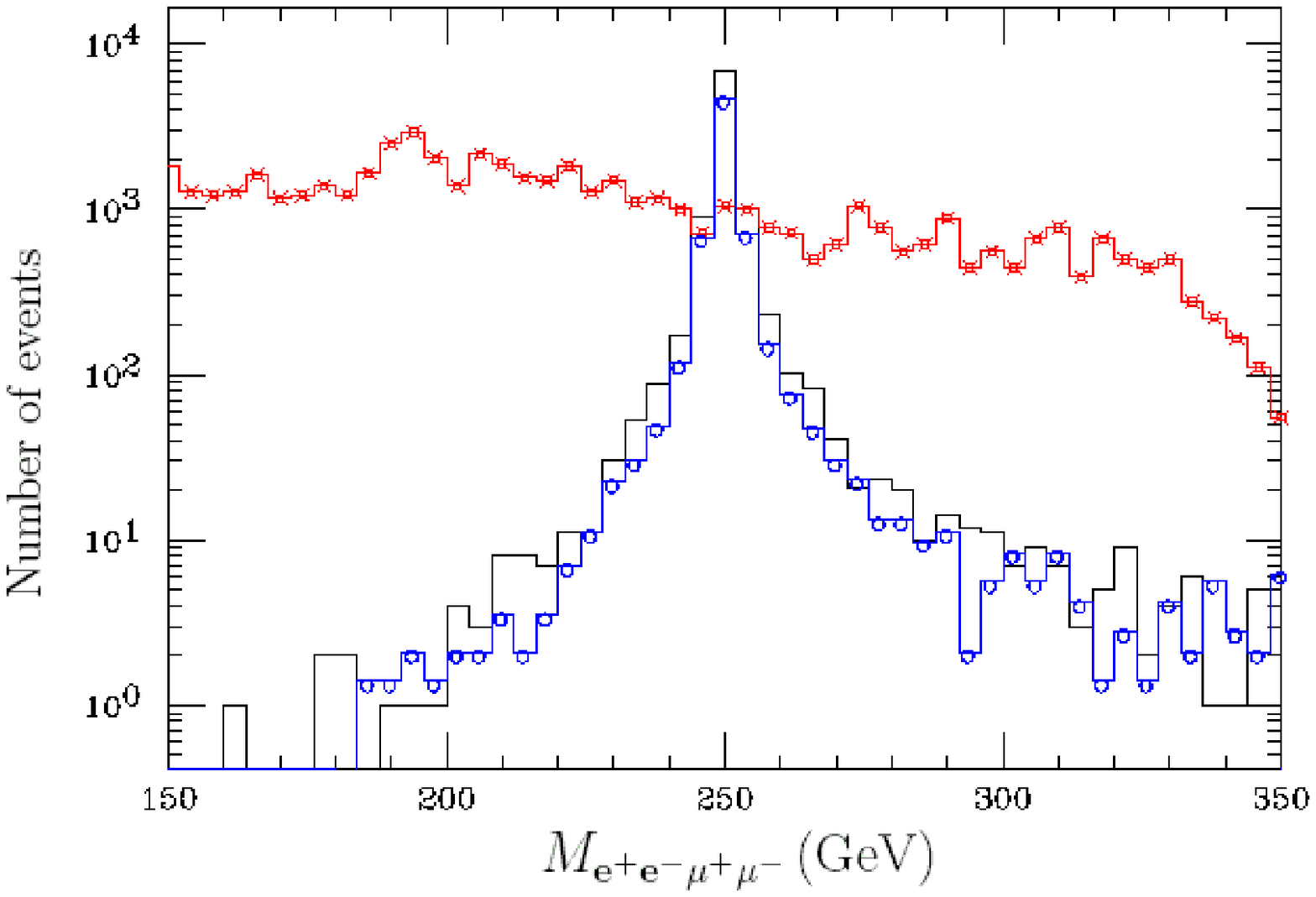}
\caption{$M_{e^+ e^- \mu^+ \mu^-}$ (in GeV) versus number of events (arbitrary luminosity) for benchmark point D at the Tevatron (top), and point F at the LHC (bottom). No cuts are applied yet. Black solid line represents $h \rightarrow XX \rightarrow 4l$ signal, red crossed $ZZ(\gamma) \rightarrow 4l$, and blue circled $h \rightarrow ZZ \rightarrow 4l$. \label{mijkl.FIG}}
\end{center}
\end{figure}

To help in distinguishing signal from background, we make various kinematical cuts.
We pair two opposite sign leptons with $\Delta R_{\ell^+ \ell^-} < 2.5$
to ensure that they come from the
same $\zpri$, and for this pair, form the dilepton invariant mass $M_{\ell^+\ell^-}$.
We also form the 4-lepton invariant mass $M_{\ell^+\ell^-\ell^+\ell^-}$. In Fig.~\ref{mijkl.FIG}, we show 4-lepton invariant mass plots for point A at LHC and point F at Tevatron, for reference.

Based on the differential distributions, we impose the following cuts
in order to maximize signal over background:
\bea
{\rm Basic\ cuts :} &\ & {p_T}_\ell \geq 20, 10, 10, 10\gev \quad ; \quad |\eta_\ell| < 2.5 \ , \nonumber \\
  \Delta R\ {\rm cut :} &\ & 0.05 < \Delta R_{\ell^+ \ell^-} < 2.5 \ , \nonumber \\
  M_{ij}\ {\rm cuts :}&\ &  M_{ee} = M_{\mu\mu} \pm 10\gev \ , \nonumber \\
  M_{ijkl}\ {\rm cut :} &\ & M_{ee\mu\mu} = M_h \pm 10\gev \quad  \ .
  \label{lepton cuts}
\eea
The four-lepton cut around ``$M_h$" is achieved by hypothesizing a Higgs boson resonance and
scanning across that hypothesis.
Such a scan is realizable in our case since the signal stands clearly
above the continuum background.
The signal and background cross-sections are shown in Table~\ref{csTeVLHC.TAB}.
We find that the 4-lepton invariant mass cut is most effective in reducing
the background.
The $S/B$ is good for all the benchmark points,
but can be improved further by the additional cut:
$M_{l^+l^-}\neq M_Z\pm 10 \, {\rm GeV}$,
which removes on-shell $Z$-bosons surviving in the data sample.

\begin{table}
\begin{tabular}{|c|c|c|c|c|c|c|}
\hline
Tevatron&
A&
B&
C&
D&
E&
F\tabularnewline
\hline
\hline
$\zpri\zpri$&
$8.8,4.3$&
$3.9,0.8$&
$4.2,2.4$&
$2.3,0.8$&
$0.05,0.02$&
$0.03,0.01$\tabularnewline
\hline
$hZZ$ (ab)&
$0.8,0$&
$1.4,0$&
$7.4,0$&
$12.8,0$&
$17,1.6$&
$21.4,1.8$\tabularnewline
\hline
$VV$ &
\multicolumn{2}{c|}{$9.7,4.3\times10^{-3}$}&
\multicolumn{2}{c|}{$9.7,3.5\times10^{-3}$}&
\multicolumn{2}{c|}{$9.7,0.01$}\tabularnewline
\hline
\end{tabular}

\begin{tabular}{|c|c|c|c|c|c|c|}
\hline
LHC&
A&
B&
C&
D&
E&
F\tabularnewline
\hline
\hline
$\zpri\zpri$&
$631,245$&
$236,44$&
$348,173$&
$212,57$&
$12,5.6$&
$6.5,2.2$\tabularnewline
\hline
$hZZ$ (ab)&
$0,0$&
$130,1.2$&
$630,2.3$&
$1280,2.5$&
$3440,850$&
$4840,846$\tabularnewline
\hline
$VV$&
\multicolumn{2}{c|}{$67,0.02$}&
\multicolumn{2}{c|}{$67,0.03$}&
\multicolumn{2}{c|}{$67,0.3$}\tabularnewline
\hline
\end{tabular}
\caption{Signal and background cross-sections in fb (only $hZZ$ in ab)
for the Tevatron and LHC
in the form:  (basic cuts, all cuts). ``Basic cuts" refers only to the $p_{T_\ell}$ and $\eta_l$ cuts
in the first line of Eq.~\ref{lepton cuts}.
$VV$ denotes the contributions from $ZZ$, $\gamma\gamma$ and $\gamma Z$.
$K$-factors have not been included.
\label{csTeVLHC.TAB}}
\end{table}


{\it Conclusions.}
In our chosen example cases with large mixing among the SM and hidden sector Higgs bosons and light-enough
$M_{Z'}$ for $h\to Z'Z'$ to be on-shell, the prospects for seeing the signal at the LHC are excellent.
The signals for the various examples are well above background after all cuts have been applied.
The Tevatron is also beginning to achieve the sensitivity required to see the
signal; however, there the key challenge  is not signal to background, but overall signal rate and
luminosity to collect enough events to reconstruct a resonance.  Once sufficient luminosity is achieved, and after more tailored techniques are applied to the problem, such as those to search for SM $ZZ$ events~\cite{Tevatron techniques}, the Tevatron should be in a position to probe well a light Higgs boson decaying in the manner proposed here.

\vspace*{0.3cm}
{\it Acknowledgments.}
We are grateful to  S. Protopopescu, J. Qian and M. Strassler for discussions,
and J. Alwall and R. Frederix for generous technical help.
This work is supported in part by the DOE. SG is supported in part by the DOE grant DE-AC02-98CH10886 (BNL). SJ is supported in part by Samsung Scholarship.





\begin{thebibliography}{99}

\bibitem{Schabinger:2005ei}
  R.~Schabinger and J.~D.~Wells,
  Phys.\ Rev.\  D {\bf 72}, 093007 (2005)
  [arXiv:hep-ph/0509209].

\bibitem{Strassler:2006im}
  M.~J.~Strassler and K.~M.~Zurek,
  Phys.\ Lett.\  B {\bf 651}, 374 (2007)
  [arXiv:hep-ph/0604261].

\bibitem{Patt:2006fw}
  B.~Patt and F.~Wilczek,
  arXiv:hep-ph/0605188.

\bibitem{Kumar:2006gm}
  J.~Kumar and J.~D.~Wells,
  Phys.\ Rev.\  D {\bf 74}, 115017 (2006)
  [arXiv:hep-ph/0606183].

\bibitem{John:today}
  J.~March-Russell, S.~M.~West, D.~Cumberbatch and D.~Hooper,
  arXiv:0801.3440.

\bibitem{Babu:1997st}
  K.~S.~Babu, C.~F.~Kolda and J.~March-Russell,
  Phys.\ Rev.\  D {\bf 57}, 6788 (1998)
  [arXiv:hep-ph/9710441].

\bibitem{Chang:2006fp}
  W.~F.~Chang, J.~N.~Ng and J.~M.~S.~Wu,
  Phys.\ Rev.\  D {\bf 74}, 095005 (2006)
  [arXiv:hep-ph/0608068] and
  Phys.\ Rev.\  D {\bf 75}, 115016 (2007)
  [arXiv:hep-ph/0701254].


\bibitem{Rizzo:2006nw}
  T.~G.~Rizzo,
  arXiv:hep-ph/0610104.

\bibitem{Langacker:2008yv}
  P.~Langacker,
  arXiv:0801.1345 [hep-ph].


\bibitem{Barbieri:2005ri}
  R.~Barbieri, T.~Gregoire and L.~J.~Hall,
  arXiv:hep-ph/0509242.

\bibitem{Chacko:2005vw}
  Z.~Chacko, Y.~Nomura, M.~Papucci and G.~Perez,
  JHEP {\bf 0601}, 126 (2006)
  [arXiv:hep-ph/0510273].

\bibitem{Bowen:2007ia}
  M.~Bowen, Y.~Cui and J.~D.~Wells,
  JHEP {\bf 0703}, 036 (2007)
  [arXiv:hep-ph/0701035].

\bibitem{Foot:1991bp}
  R.~Foot, H.~Lew and R.~R.~Volkas,
  Phys.\ Lett.\  B {\bf 272}, 67 (1991);
  K.~S.~Babu and G.~Seidl,
  Phys.\ Rev.\  D {\bf 70}, 113014 (2004)
  [arXiv:hep-ph/0405197];
  R.~Foot and X.~G.~He,
  Phys.\ Lett.\  B {\bf 267}, 509 (1991).


\bibitem{Strassler:2008bv}
$H\to XX$ has been noted in a related context by
  M.~J.~Strassler,
  arXiv:0801.0629.

\bibitem{Holdom:1985ag}
  B.~Holdom,
  Phys.\ Lett.\  B {\bf 166}, 196 (1986).

\bibitem{Yao:2006px}
  W.~M.~Yao {\it et al.}  [Particle Data Group],
  J.\ Phys.\ G {\bf 33}, 1 (2006).

\bibitem{Alcaraz:2006mx}
  J.~Alcaraz {\it et al.}  [ALEPH Collaboration],
  arXiv:hep-ex/0612034.

\bibitem{Peskin:2001rw}
  M.~E.~Peskin and J.~D.~Wells,
  Phys.\ Rev.\  D {\bf 64}, 093003 (2001)
  [arXiv:hep-ph/0101342].

\bibitem{Djouadi:1997yw}
  A.~Djouadi, J.~Kalinowski and M.~Spira,
  Comput.\ Phys.\ Commun.\  {\bf 108}, 56 (1998)
  [arXiv:hep-ph/9704448];
  Program can be found at:
  http://people.web.psi.ch/spira/proglist.html .

\bibitem{Spira:1995mt}
  M.~Spira,
  arXiv:hep-ph/9510347.
  Program can also be found at http://people.web.psi.ch/spira/proglist.html .

\bibitem{Djouadi:2005gi}
  A.~Djouadi,
  arXiv:hep-ph/0503172.

\bibitem{atlas tdr}
  ATLAS Technical Design Report, Vol. 2, CERN/LHCC/99-14
  (1999), secs.\ 19.2.5 \& 19.2.9.

\bibitem{Stelzer:1994ta}
  T.~Stelzer and W.~F.~Long,
  Comput.\ Phys.\ Commun.\  {\bf 81}, 357 (1994)
  [arXiv:hep-ph/9401258].
  Program can be found at http://madgraph.phys.ucl.ac.be/index.html .

\bibitem{Maltoni:2002qb}
  F.~Maltoni and T.~Stelzer,
  JHEP {\bf 0302}, 027 (2003)
  [arXiv:hep-ph/0208156].
  Program can also be found at http://madgraph.phys.ucl.ac.be/index.html .

\bibitem{Pumplin:2005rh}
  J.~Pumplin, A.~Belyaev, J.~Huston, D.~Stump and W.~K.~Tung,
  JHEP {\bf 0602}, 032 (2006)
  [arXiv:hep-ph/0512167].

\bibitem{Tevatron techniques}
D0 Collaboration, Note 5345-CONF (March 9, 2007); CDF Collaboration, Note 8775 (5 April 2007).

\bibitem{third possibility}
A third renormalizable operator involves a SM gauge singlet fermion $N$ with the
Yukawa interaction $NLH$ related to neutrino mass generation (note however that
$N$ carries lepton number). We will not discuss this further in this work since
our discussion here is largely independent of the presence of such an interaction.

\end{thebibliography}
\end{document}